\documentclass[final,12pt]{elsarticle}

\usepackage{amssymb}
\usepackage{caption,subcaption}
\usepackage{graphicx,lineno,amsmath}

\journal{arXiv}

\begin{document}

\begin{frontmatter}



\title{Impact of photon cross section systematic uncertainties on Monte Carlo-determined depth-dose distributions}


\author{E. Aguirre}
\author{M. A. Bernal}
\ead{mbernalrod@gmail.com}

\address{Instituto de F\'isica Gleb Wataghin. Universidade Estadual de Campinas. Brazil.}
\author{M. David, C. E. deAlmeida}
\address{Laborat\'orio de Ci\^encias Radiol\'ogicas. Universidade do Estado do Rio de Janeiro. Brazil.}%

\begin{abstract}
This work studies the impact of systematic uncertainties associated to interaction cross sections on depth dose curves determined by Monte Carlo simulations. The corresponding sensitivity factors are quantified by changing cross sections in a given amount and determining the variation in the dose. The influence of total cross sections for all particles, photons and only for Compton scattering is addressed. The PENELOPE code was used in all simulations. It was found that photon cross section sensitivity factors depend on depth. In addition, they are positive and negative for depths below and above an equilibrium depth, respectively. At this depth, sensitivity factors are null. The equilibrium depths found in this work agree very well with the mean free path of the corresponding incident photon energy. Using the sensitivity factors reported here, it is possible to estimate the impact of photon cross section uncertainties on the uncertainty of Monte Carlo-determined depth dose curves.
\end{abstract}


\end{frontmatter}

\section{\label{sec:level1}Introduction}

The determination of any physical quantity carries an associated uncertainty, even when it is not intrinsically stochastic. The radiation-matter interaction is a stochastic process so any quantity used to describe it, would have an intrinsic stochastic behavior. Thus, the Monte Carlo (MC) method emerges as a natural tool to study this process. 

Commonly, a measurand  is not directly measured. Instead, it can be determined  from other quantities. That is, $y=f(X_1, X_2, ...,X_N)$, where $y$ is the measurand and $X_i$ is set of input quantities. According to the {\it Bureau International des Poids et Mesures} (BIPM)\cite{BIPM_uncertainty08}, the uncertainty of some input quantities can be determined during the measurement and the others carry an associated uncertainty determined previously. The former uncertainties can be determined, for instance, by repeated observations or by judgements based on experience. The latter may be extracted from calibration certificates and reference data. In addition, uncertainties can be classified according to the method of evaluation. Type A evaluation of a standard uncertainty can be carried out after having collected $n$ independent observations
\begin{equation}
{\bar x_i}={\i\over n}\sum^n_{k=1}x_{i,k}
\end{equation}

This arithmetic mean is used as the input estimate for the function $y$, that is, $X_i={\bar x_i}$. This quantity fluctuates due to random variation of the influence quantities during measurements or due to its intrinsic stochasticity. Such a fluctuation can be quantified through the variance of the observations

\begin{equation}
s^2(x_i)={1\over n-1}\sum^n_{k=1}(x_{i,k}-{\bar x_i})^2
\end{equation}

This variance represents the variability of $x_i$ around its mean value ${\bar x_i}$. Finally, the uncertainty associated to the input variable, which is the mean value of $x_i$, can be estimated as

 \begin{equation}
s^2({\bar x_i})={s^2(x_i)\over n}.
\end{equation}

$s^2$ is a measure of the uncertainty of ${\bar x_i}$ and quantifies how good ${\bar x_i}$ estimates the real expectation value of $x_i$, which is most of the time inaccessible to us.  Thus, $s^2({\bar x_i})$ and $s({\bar x_i})$ are the Type A variance and Type A standard uncertainty of the input quantity $X_i$, respectively.

If $x_i$ is a previously determined quantity, its associated uncertainty can be estimated from the available information on its variability. For instance, from previous experimental data, calibration certificate, data uncertainties reported in handbooks, and so forth. This is the so called Type B evaluation of standard uncertainty. For sake of simplicity, these uncertainties are referred to as Type A or statistical and Type B or systematic uncertainties.

When the input variables are  independent or uncorrelated, the  combined variance $s^2(y)$ is given by

\begin{equation}
s^2(y)=\sum^N_{i=1}{\partial f\overwithdelims () \partial x_i}^2s^2({\bar x_i}).
\label{propagation}
\end{equation}

The term between parenthesis in eq. \ref{propagation} is known as the sensitivity  factor, which is a measure of how sensitive is the quantity $y$ with respect to variable $x_i$. Variances $s^2({\bar x_i})$ may be estimated according to the Type A or Type B evaluation methods, depending on the problem in question.

MC simulations have been widely used in Medical Physics during the last two decades \cite{Verhaegen99,Rogers06}. It has been even used to calculate physical quantities that are very difficult to determine experimentally \cite{Sanchez07}. Based on the fact that this approach is used to simulate stochastic phenomena, quantities determined in these simulations have an intrinsic statistic uncertainty. This uncertainty is the only one reported in the vast majority of scientific articles in which MC simulations were used. Several works have addressed the impact of cross section uncertainties in the results of MC simulations, yet not all in a systematic way. Demarco et al. \cite{Demarco02} used the MCNP4C MC code \cite{Breismeister93} to study the impact of different photon cross section sets on absorbed dose in water in the 10-1000 keV energy range. They compared the cross section libraries DLC-015 \cite{Storm70}, DLC-146 \cite{Saloman88} and DLC-200 (default of MCNP4C) to the XCOM compilation from  the National Institute of Standard and Technology (NIST) \cite{Berger99}. The DLC-200 tabulation departs the most from the XCOM set. At 40 keV, there is an extreme underestimation of the absorbed dose at 1 cm from a point source determined with the DLC-200 set when compared to that obtained with the XCOM compilation. This seems to be caused by a corresponding underestimation of the photoelectric cross section (DLC-200.vs.XCOM), which is of about 10\%. Below 10 keV, the ratio between the DLC-200 and XCOM absorbed doses tends to diverge through an overestimation. A year later, Bohm et al. \cite{Bohm03} benchmarked the MCNP/MCNPX MC code (see Ref. \cite{Goorley14} and references there in) for the characterization of low energy brachytherapy sources (also with $^{125}$I and $^{103}$Pd). They carried out calculations using two cross section sets and reported on the consequent impact on  the corresponding dose rate constants and radial dose functions. According to their results,  differences of about 10\% and 6\%  in the photoelectric and total cross sections in this energy range, respectively, lead to changes in the dose rate constant of 3\% and 5\%, and of 18\% and 20\% in the radial dose function for  the $^{125}$I and $^{103}$Pd sources, respectively. Williamson and Rivard \cite{Williamson09} carried out a more systematic work on uncertainty propagation when determining brachytherapy sources dosimetric parameters via Monte Carlo simulations. They proposed the following formula to account for systematic and statistical uncertainties of a given quantity (Y):
\begin{eqnarray}
    \%\sigma_Y&=&\sqrt{\%\sigma^2_{Y|\mu}+\%\sigma^2_{Y|geo}+\%\sigma^2_{Y|s}}\\
    &=&\sqrt{\bigg (\%{\partial Y\over \partial\mu}\bigg )^2\%\sigma_\mu^2+\bigg (\%{\partial Y\over \partial geo}\bigg )^2\%\sigma_{geo}^2+\bigg (\%{\partial Y\over \partial s}\bigg )^2\%\sigma_s^2},
\end{eqnarray}

where $\mu$ is the total attenuation coefficient, $geo$ is a geometrical parameter and $s$ is the statistical uncertainty associated to the MC simulation. $Y$ can be any of the interest quantities, such as the dose rate constant ($\Lambda$), the radial dose function ($g(r)$), among others. They used two cross section sets to estimate the corresponding sensitivity factor and obtained, for instance, $\%\partial\Lambda/\partial\mu=0.68$.

Rogers and Kawrakow \cite{Rogers03} published a work tackling the systematic uncertainty problem in MC simulations in a more consistent way. Specifically, they studied the sensitivity of several correction factors used in the Canadian air-kerma primary standard, to some influence parameters. These parameters were the MC code and transport algorithm, the $^{60}$Co spectrum, the source diameter, and the source-chamber distance. More recently, Wulff et al. \cite{Wulff10} studied the systematic uncertainties of ionization chamber quality correction factors ($k_Q$) determined by MC simulations. Various uncertainty sources were investigated, including geometrical factors and interaction cross sections. For photons, they scaled the corresponding cross section whereas for electrons, the mean excitation energy was varied as it is the main source of uncertainty for stopping powers. This way, sensitivity factors were found and combined with the corresponding estimated uncertainty to know how cross section uncertainties propagate to $k_Q$.  In a similar work, Muir and Rogers \cite{Muir10} carried out a very detailed analysis on systematic uncertainty sources when determining the beam quality correction factors for ionization chambers (k$_Q$). They included items such as photon cross sections, electron stopping power, chamber dimensions, and photon spectra. 

Recently, Ali et al. \cite{Ali15} used a very sophisticated statistical method to derive uncertainties of photon cross sections. This method was firstly applied to experimental and Monte Carlo simulated transmission factors for several megavoltage beams. Secondly, the approach was directly applied to experimental and theoretical (XCOM+IAEA set) photon interaction cross sections. Only graphite and lead materials were included in this study. Unfortunately, this method was not sensitive enough to resolve the energy dependence of the uncertainties in question. Thus, the authors reported energy-independent photon cross section uncertainties of 0.6\% and 0.2\% for graphite and lead, respectively (68\% confidence), obtained from the comparison between experimental and simulated transmission curves. These uncertainties were 0.2\% and 0.9\% (68\% confidence) for graphite and lead, respectively, when the method was directly applied to experimental and theoretical cross sections. For practical reasons, the authors recommend use an overall CS uncertainty of 0.5\% (68\% confidence) for photons with energies from 100 keV up to 40 MeV. 

The uncertainties reported by Ali et al. \cite{Ali15} are lower than those previously estimated by the seminal compilation of Hubbell \cite{Hubbel99}. According to this review, photoeffect cross section uncertainties can be estimated as  2\% and 1\%-2\% (68\% confidence) for photons with energies of 5 keV-100 keV and 100 keV-10 MeV, respectively. In addition, Hubbell recommends an overall uncertainty for the total mass attenuation coefficient ($\mu/\rho$) of the order of 1\% to 2\%  (68\% confidence) for photon energies from 5 keV up to a few MeV.

One of the most important source of systematic uncertainty in MC simulations is interaction cross section. Besides the approach followed by Ali et al. \cite{Ali15}, there are two other approaches for evaluating the impact of systematic uncertainties on a quantity determined by MC simulations. The first one could be letting the input quantity fluctuate during simulation according to a given distribution and the corresponding variance. Namely, this quantity can be sampled during each MC step. A second simulation can be done by choosing the expected value of the same quantity, or its reference value, during each MC step. That is, not letting it fluctuate so the final result will only represent a pure statistical uncertainty. Then, the two corresponding final uncertainties can be compared and the influence of the systematic uncertainty on the quantity in question can be deconvoluted. This method could be regarded as a statistical or Type A evaluation of the uncertainty associated to a systematic effect. That is, the systematic effect of this quantity is converted into a random effect during its evaluation (see $\S$3.3.3 of \cite{BIPM_uncertainty08}). The second one is simpler. The input quantity can be artificially scaled, let's say in $\pm$5 \%. A long enough MC simulation would give back a negligible statistic uncertainty, then the sensitivity factor $\partial f/ \partial x_i$ can be determined. This factor can be used in conjunction with the known systematic uncertainty of $x_i$  to calculate the  contribution of this variable to the quantity $y$ uncertainty. The latter approach was followed in this work.

A complete study of systematic uncertainties in MC simulations for the determination of absorbed dose distributions would involve several parameters, such as both differential and total interaction cross sections, material composition and density, geometrical factors, and so forth. The systematic study of electron CS must include all the transport parameters related to the history condensation algorithm. This algorithm and the associated controlling parameters are particular for each MC code. This could be the matter for another independent work. For these reasons, electron CS uncertainties were left out of the scope of this work. In this study, only the influence of total photon cross sections on the central depth dose distribution of a primary photon beam is investigated. This influence is quantified trough the determination of the dose-to-CS sensitivity factors. The methodology used in this work is able to account for energy dependency of the dose-to-CS sensitivity factors. That is, the impact of photon CS uncertainty  on dose uncertainty can be energy-resolved. The sensitivity factors reported here can be used in conjunction with CS uncertainties to quantified the corresponding systematic uncertainty of MC-determined absorbed dose. The basic geometry in our work is a photon beam impinging a water phantom. Broad scattering conditions, as that found in brachytherapy applications, are out of the scope of this work.

\section{Methods}
\subsection{The  PENELOPE code }
PENELOPE is a Monte Carlo code for simulating coupled photon-electron transport in virtually any medium \cite{PENELOPE08}. This code can handle the transport of photon, electrons and positrons from about 1 GeV down to 50 eV. A mixed simulation strategy can be used to simulate electron and positron transport, in which part of the history of the charged particle is condensed and the other is treated in detail. In addition, the user can set the code up to follow every charged particle step-by-step, namely, in an analog simulation. PENELOPE is a well documented MC code with a relatively simple structure, which makes it suitable for our purposes in this study. On the one hand, C$_1$ and C$_2$ parameters control the mean free path between  hard or catastrophic interactions in the mixed simulation algorithm. On the other hand, W$_{cr}$ and W$_{cc}$ represent the threshold energy for the creation of a secondary particle through radiative and collisional events, respectively. As in all MC code, photon (P$_{abs}$) and electron (E$_{abs}$) cutoff transport energies have to be defined.

\subsection{Simulations settings}
The user code PENMAIN was used in our simulations. In this case, the geometry is defined through an input file with extension .geo. The simulation geometry consists of a homogeneous water phantom with density of 1.0 g/cm$^3$.The front face of the phantom is covered with an air slab. Photons were the primary particles in all simulations, with energies of 20 keV, 100 keV, 1.25 MeV, 2 MeV, and 5.0 MeV. Interaction cross sections (CS) were pre-calculated with the accompanying code material.exe. Pencil beams were used for all photon energies. In addition,  3x3 cm$^2$ normal and divergent beams were simulated only for $^{60}$Co photons in order to explore possible  effects of the extension and divergence of the beam on the sensitivity factors.
The dimensions of the water phantom was 10x10x5 cm$^3$ for 20 keV photons, 10x10x15 cm$^3$ for 100 keV photons, 10x10x25 cm$^3$ for $^{60}$Co and 2 MeV, and 10x10x35 cm$^3$ for 5 MeV photons. All phantoms were divided into 20x20x50 uniform bins to define the voxels for dose distribution determination. The transport parameters defined above were set as follows. C$_1$=C$_2$=0.01, W$_{cc}$=W$_{cr}$=100 eV, and E$_{abs}$=P$_{abs}$=1 keV for 20 keV electrons. C$_1$=C$_2$=0.01, W$_{cc}$=W$_{cr}$=1 keV, and E$_{abs}$=P$_{abs}$=10 keV for the other photon energies.

\subsection{Systematic uncertainty evaluation.}

Total (not differential) CS for photons were artificially varied in $\pm5$ and $\pm10$ \% by directly scaling the inverse mean free path in the same proportion during simulations. This scaling factor was denoted as $f$. This was done by modifying the penelope.f code. The influence of CS uncertainties on dose distributions was study for the four most important interaction events for the energy range in question, namely Rayleigh and Compton scattering, photo-electric effect and pair production. As the importance of each event type depends on the photon energy, not all the mentioned events were studied for each incident energy. CS for Rayleigh and Compton scattering and photo-electric effect were scaled for 20 keV photons. Compton scattering and photo-electric effect CS were scaled for 100 keV. For $^{60}$Co photons, only Compton CS was included in calculations. Compton scattering and pair-production CS were studied for 2 MeV and 5 MeV photons. 

The ratio (R) between the doses with and without CS scaling was determined. Long enough simulations were carried out to obtain statistic uncertainties sufficiently low to resolve the influence of the systematic uncertainty on depth dose curves. The uncertainty associated to the dose was determined according to the following approach. If only the uncertainty  due to a given physical process cross section ($s_{cs}$) is accounted for, then the dose uncertainty ($s_D$) is
\begin{equation}
s_D={\partial D\over \partial cs} s_{cs}.
\label{dose_cs1}
\end{equation}

If the right-hand side of eq. \ref{dose_cs1} is divided and multiplied by the unscaled CS ($cs_0$), we get

\begin{equation}
s_D={\partial D\over \partial f}{s_{cs}\overwithdelims () cs_0},
\label{dose_cs2}
\end{equation}

where $f$ is the CS scaling factor and the last term between parenthesis is the CS relative standard deviation. This equation can be manipulated to obtain
\begin{equation}
{s_D\over D_0}={\partial R\over \partial f}{s_{cs}\overwithdelims () cs_0},
\label{dose_cs3}
\end{equation}

where $R$ is the relative dose, as defined above. Thus, the sensitivity factors $\partial R/\partial f$ were determined in the current work by scaling CS and determining the function $R.vs.f$. These factors can be used in conjunction with the relative standard deviation of CS  to propagate their  uncertainties to the dose. The left-hand side of eq. (\ref{dose_cs3}) represents the relative dose uncertainty due to the CS uncertainty.

\section{RESULTS AND DISCUSSION.}

\begin{figure}
    \begin{subfigure}[b]{0.45\textwidth}
        \includegraphics[width=\textwidth]{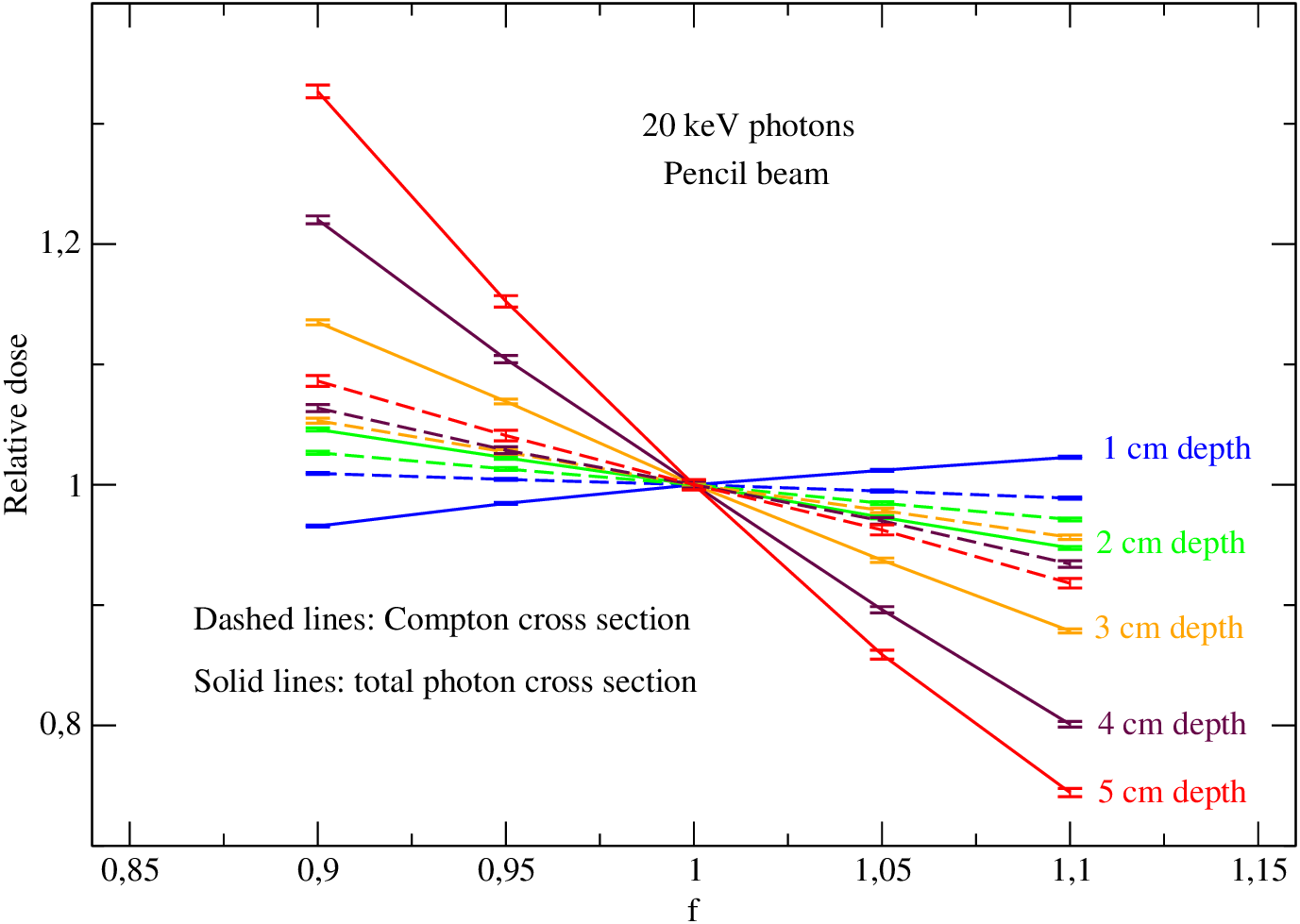}
        \caption{Compton}
       \label{fig_20keV_a}
    \end{subfigure}
    \hspace{1cm}
    \begin{subfigure}[b]{0.45\textwidth}
        \includegraphics[width=\textwidth]{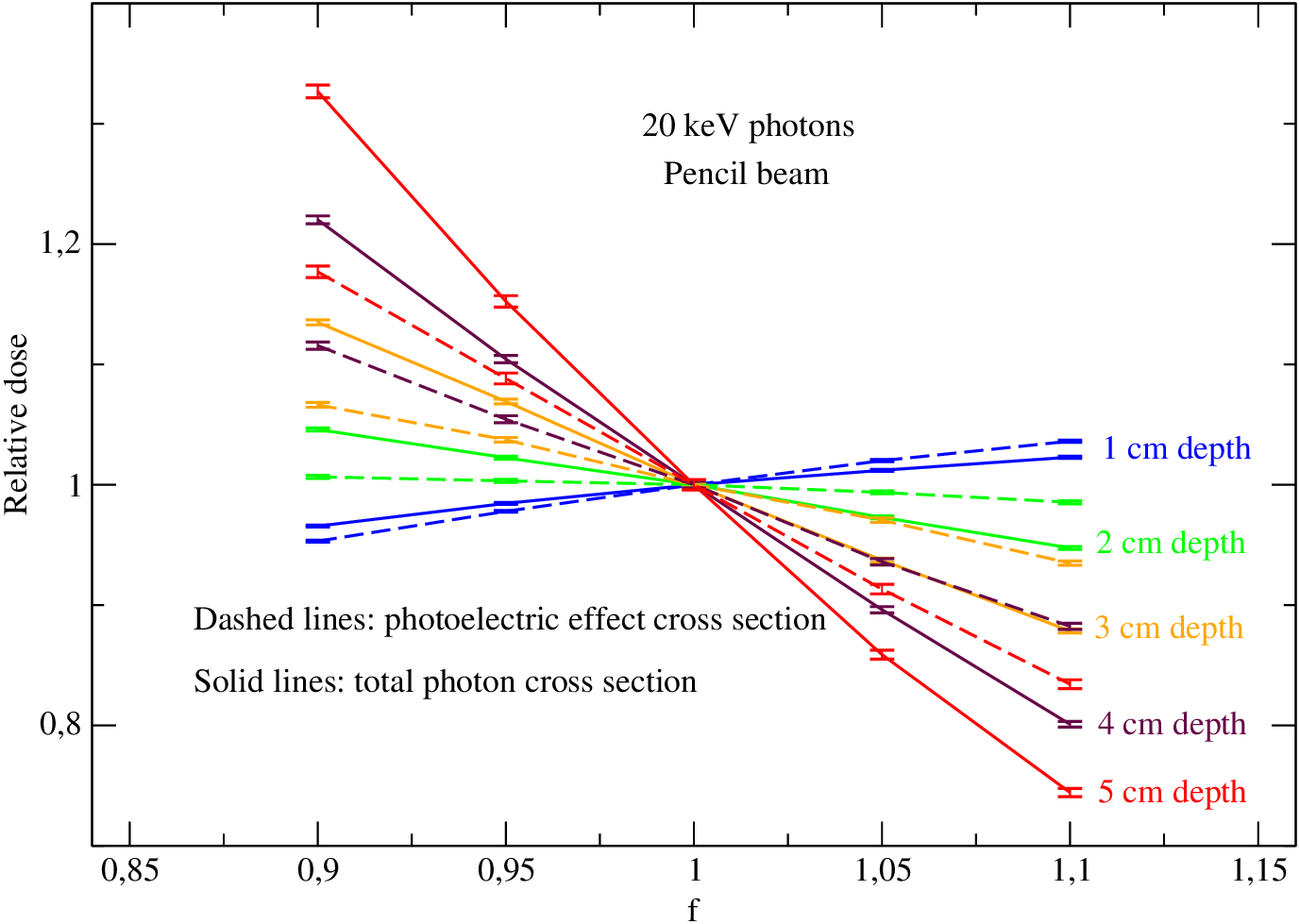}
         \caption{Photo-electric}
         \label{fig_20keV_b}
    \end{subfigure}
   \\
   
   \centering
    \vspace{1cm}
     \begin{subfigure}[b]{0.45\textwidth}
        \includegraphics[width=\textwidth]{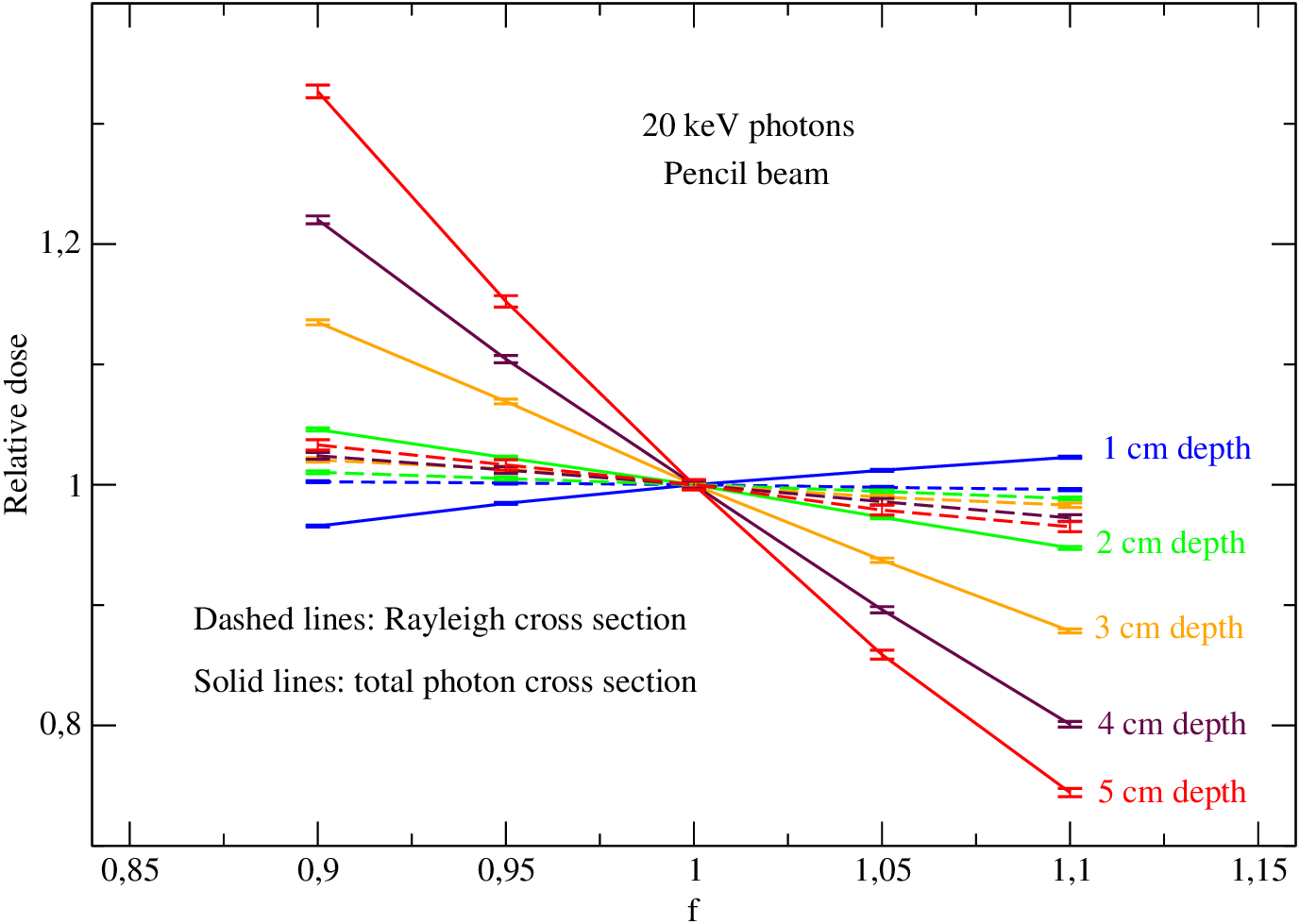}
         \caption{Rayleigh}
         \label{fig_20keV_c}
    \end{subfigure}
\caption{\label{fig_20keV}Ratio between the central axis doses for a 20 keV pencil beam with and without scaling of total cross sections as a function of the scaling factor $f$. Solid lines represent the situation in which total cross sections were scaled. Dashed lines represent the cases where (a) Compton, (b) photo-electric, and (c) Rayleigh CS were scaled.  Values are reported for five different depths.}

\end{figure}

Figure \ref{fig_20keV} shows the ratio between the central axis doses obtained with and without CS scaling as a function of the scaling factor $f$ for five depths. These ratios were obtained by scaling total and only Compton CS. In this case, a 20 keV pencil beam was simulated. The first important feature observed in this figure is that the gradient $\partial R/\partial f$ is positive for depths below a certain value, which will be from now on denoted as the equilibrium depth (ED). Above the ED, this gradient becomes negative. In other words, the dose increases with the increment of CS at depths lower than the ED. However, this behavior is inverted for depths greater than the ED. This can be explained as follows. The mean free path ($\lambda$) of 20 keV  photons in liquid water is about 1.24 cm (determined from data found at Ref. \cite{NISTphotons}), which seems to be very close to the ED extracted from these results (1.3 cm).  Thus, $\lambda$ would represent a pivot depth around which the depth dose curve turns on the dose-depth plane when CS are scaled. When CS are reduced ($f<1$), $\lambda$ increases so photons become more penetrating, carrying out their energy towards greater depths. This is why the dose is reduced at depths lower than the ED ($R<1$) and increased ($R>1$) at depths higher than the ED. On the one hand, underestimation of CS leads to the decrease (increase) of the dose at lower (greater) depths. On the other hand, overestimation of CS leads to the increase (decrease) of the dose at lower (greater) depths. In addition, the impact of CS uncertainties on the dose uncertainty is negligible at the ED because $\partial R/\partial f\approx0$. As the energy deposition process depends on the interplay between photon energy fluence ($\Psi$) and mass energy absorption coefficient ($\mu_{en}/\rho$) at a given depth, the behavior discussed just above can be also explained analyzing these quantities. When total CS is reduced, $\mu_{en}/\rho$ decreases proportionally and the energy fluence at lower depths changes little  which leads to a dose decrease. The diminution of CS makes the beam more penetrating, producing an increase of the fluence at greater depths.  Due to the exponential behavior of fluence as a function of the attenuation coefficient, this fluence should increase more than the fraction used to scale this coefficient (or CS). Then, the product $\Psi\mu_{en}/\rho$ (dose) at high depths tends to increase when scaling down CS.  At the equilibrium depth, the decrease of $\mu_{en}/\rho$ is compensated by the fluence increase so the sensitivity factor is nearly zero. Fig. \ref{fig_20keV_a} shows that the effect of scaling Compton CS is relatively weak for this photon energy. It should be noticed that the curves corresponding to Compton CS are collapsed around  $\partial R/\partial f\sim 0$. However, scaling photo-electric CS (see fig. \ref{fig_20keV_b})  has a stronger influence on the dose distribution when compared to that of the Compton scattering. This is an expected behavior since Compton scattering CS are lower than those for photo-electric effect in water below $\sim$ 26 keV. In addition, there is a peculiar behavior in the sensitivity factor at 1 cm depth. The dose at this depth is more sensitive to the scaling of photo-electric CS than to the total CS. This suggests that the effect of CS scaling in photo-electric effect is opposite to that for Compton and Rayleigh scattering CS. This could be related to the trade-off between attenuation and scattering of the photon beam. Fig. \ref{fig_20keV_c} depicts the results of scaling the Rayleigh CS. It can be observed that the effect of this scattering is even weaker than that seen for Compton scattering. This effect on the depth-dose curve is not null because, despite that this is an elastic scattering of photons with negligible energy deposition, Rayleigh scattering may remove photons from the pencil beam so the dose in the field  central axis may be affected. Notice that the sensitivity factor ins this case are always positive. If the Rayleigh scattering CS is reduced, less photons would be scattered out the beam and the central axis dose should increase. In addition, the effect on the dose distribution increases with depth since photon multiple scattering probability also increases. 

Since each depth shows a unique and almost constant gradient, the impact of a given CS uncertainty on the depth dose distribution depends on depth. That is, MC-determined doses for a target sited at depths below or above the equilibrium depth would have greater associated uncertainties than that placed near the ED.

\begin{figure}
    \begin{subfigure}[b]{0.45\textwidth}
        \includegraphics[width=\textwidth]{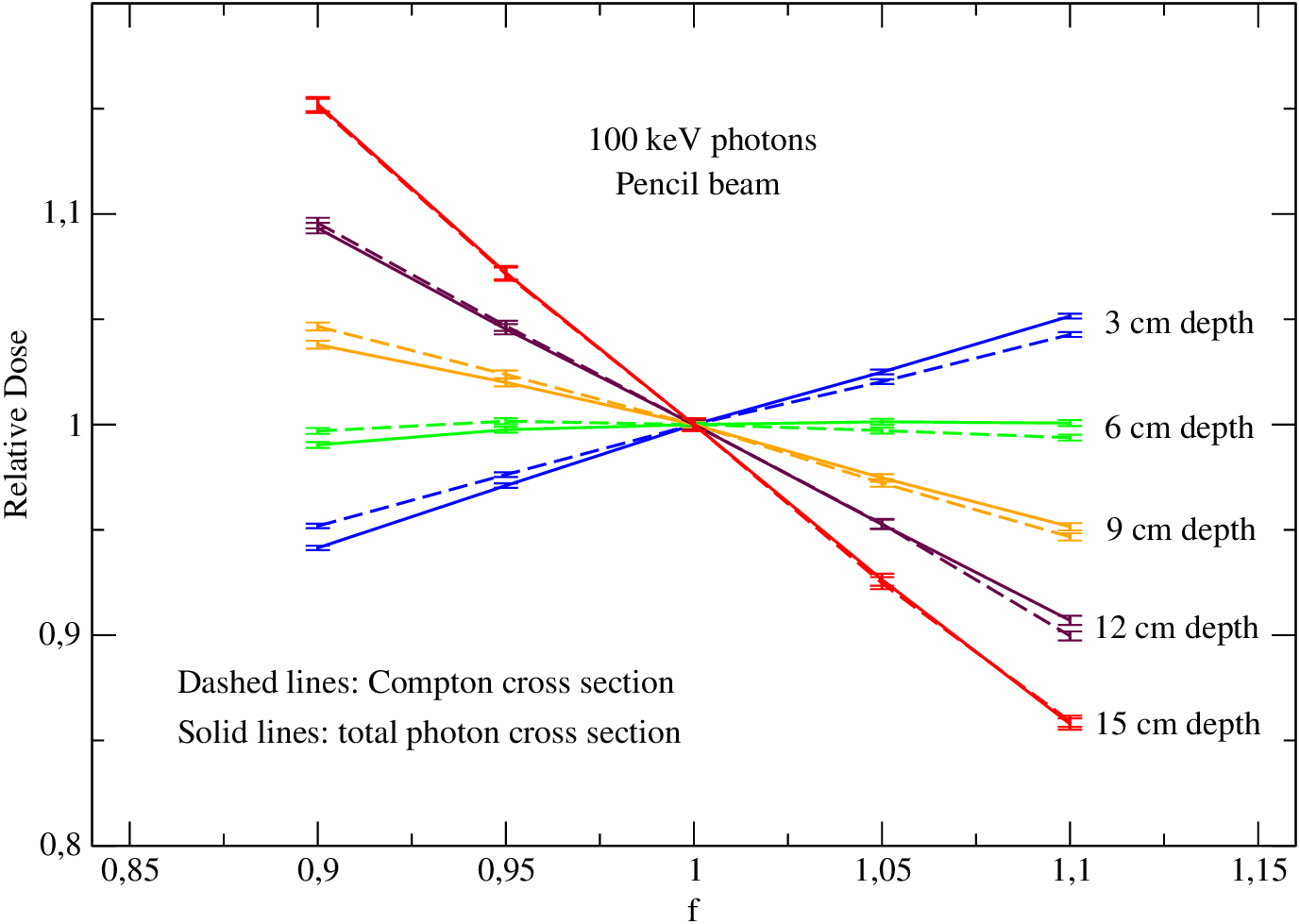}
        \caption{Compton}
       \label{fig_100keV_a}
    \end{subfigure}
    \hspace{1cm}
    \begin{subfigure}[b]{0.45\textwidth}
        \includegraphics[width=\textwidth]{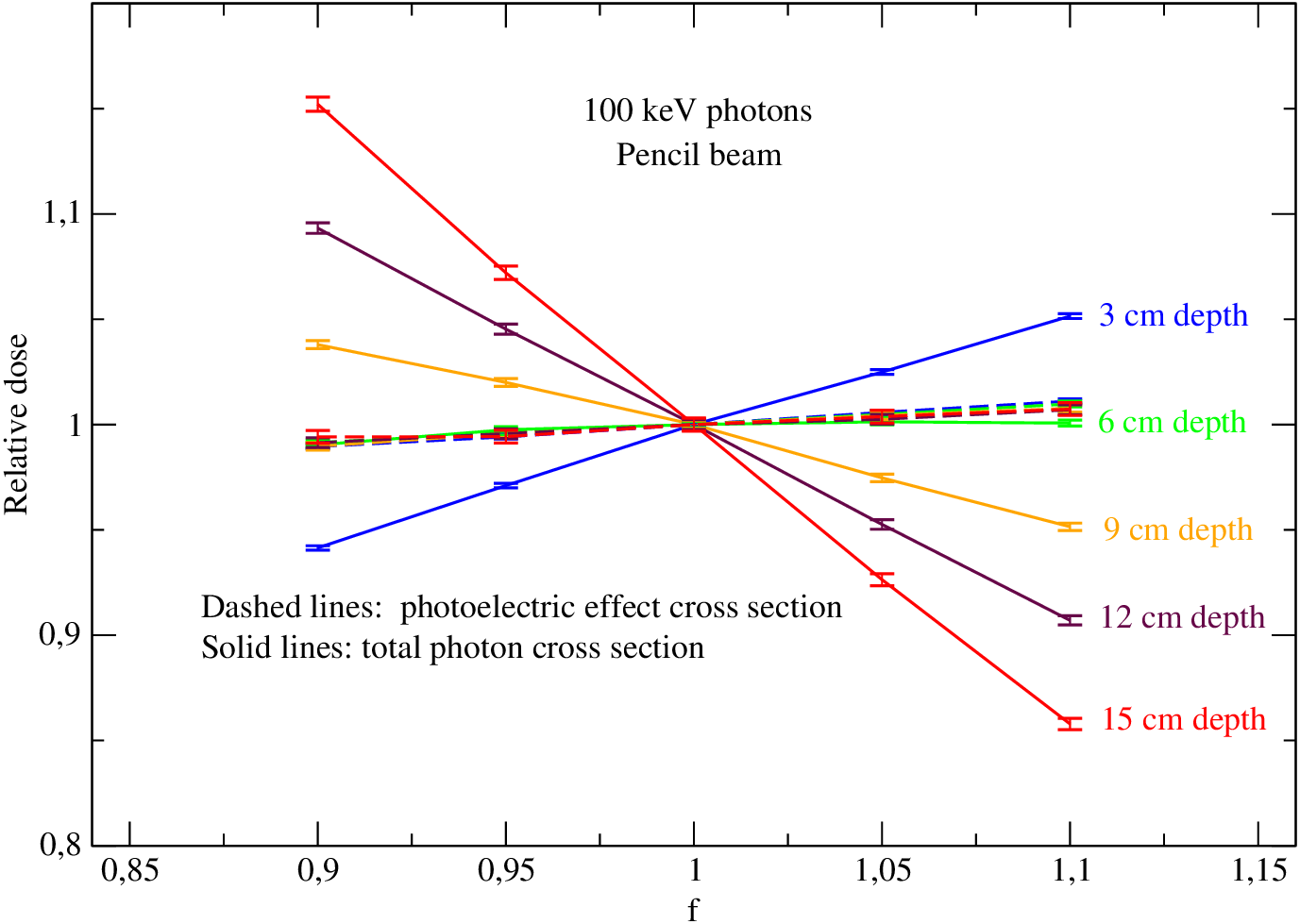}
         \caption{Photo-electric}
         \label{fig_100keV_b}
    \end{subfigure}
\caption{\label{fig_100keV}Ratio between the central axis doses for a 100 keV pencil beam with and without scaling of total cross sections as a function of the scaling factor $f$. Solid lines represent the situation in which total cross sections were scaled. Dashed lines represent the cases where (a) Compton, (b) photo-electric CS were scaled.  Values are reported for five different depths.}
\end{figure}

Fig. \ref{fig_100keV} depicts similar results as those shown in Fig. \ref{fig_20keV}, but for 100 keV photons. It is observed the same behavior explained for 20 keV photons. Namely, there is an equilibrium depth. The mean free path for 100 keV photons is about 5.9 cm \cite{NISTphotons}, which agrees very well with the ED extracted from simulations (6.0 cm). In addition, it is evident that Compton scattering is the main responsible for the sensitivity of dose to photon CS scaling at this energy. Fig. \ref{fig_100keV_b} clearly shows that the influence of the photo-electric effect is almost negligible is this case.  The influence of Rayleigh scattering was also studied but the resulting sensitivity factors are practically zero, so the  plot was omitted from this figure. Since each depth shows a different and almost constant gradient or sensitivity factor, the impact of a given CS uncertainty on the depth dose distribution depends on the depth. That is, MC-determined doses for a target sited at depths shallower or deeper than the equilibrium depth would have greater associated uncertainties than those placed near the ED.

\begin{figure}
\centering
\includegraphics[width=.5\linewidth]{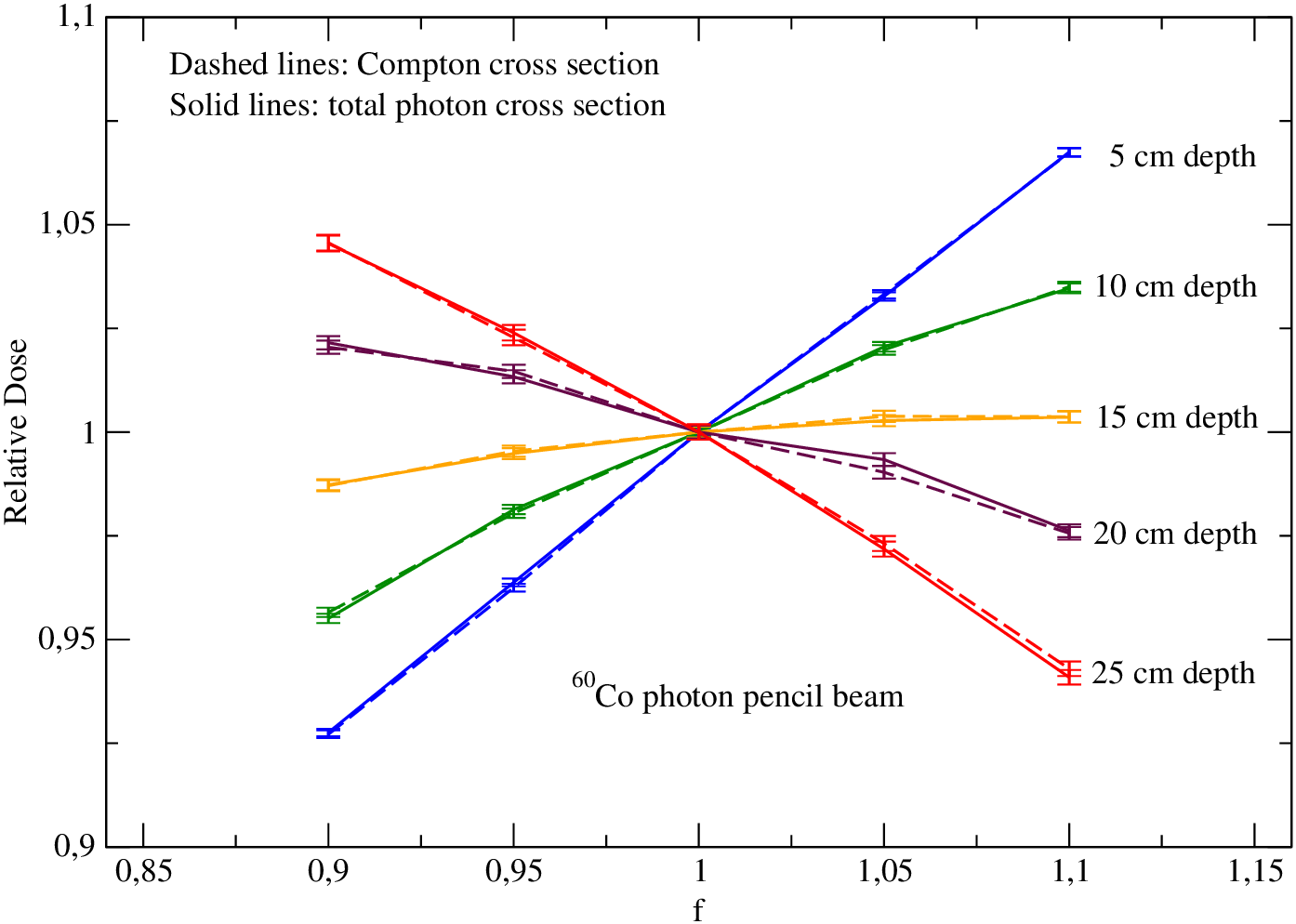}
\caption{Ratio between the central axis doses for a $^{60}$Co pencil beam with and without scaling of cross sections as a function of the scaling factor $f$. Solid and dashed lines represent the situation in which total and Compton cross sections were scaled, respectively.  Values are reported for five different depths.}
\label{fig_Co}
\end{figure}

Similar calculations were carried out for $^{60}$Co photons (see fig. \ref{fig_Co}). At this energy, the influence of Compton scattering CS scaling is overwhelming. For this reason, the influence of the photo-electric effect, Rayleigh scattering, and pair-production are not explicitly reported in this work. The mean free path of $^{60}$Co photons is about 15.8 cm \cite{NISTphotons}, which is quite similar to the one obtained in our simulations (16.0 cm). At this particular energy, the points at 5 cm and 25 cm depth are almost equidistant to the ED. However,  fig. \ref{fig_Co} shows that the sensitivity factor for the greater depth is lower than that for the other. Probably, the greater depth is less sensitive to CS scaling because at greater depth the effect of scattering may tend to compensate the influence of CS scaling.

\begin{figure}
    \begin{subfigure}[b]{0.45\textwidth}
        \includegraphics[width=\textwidth]{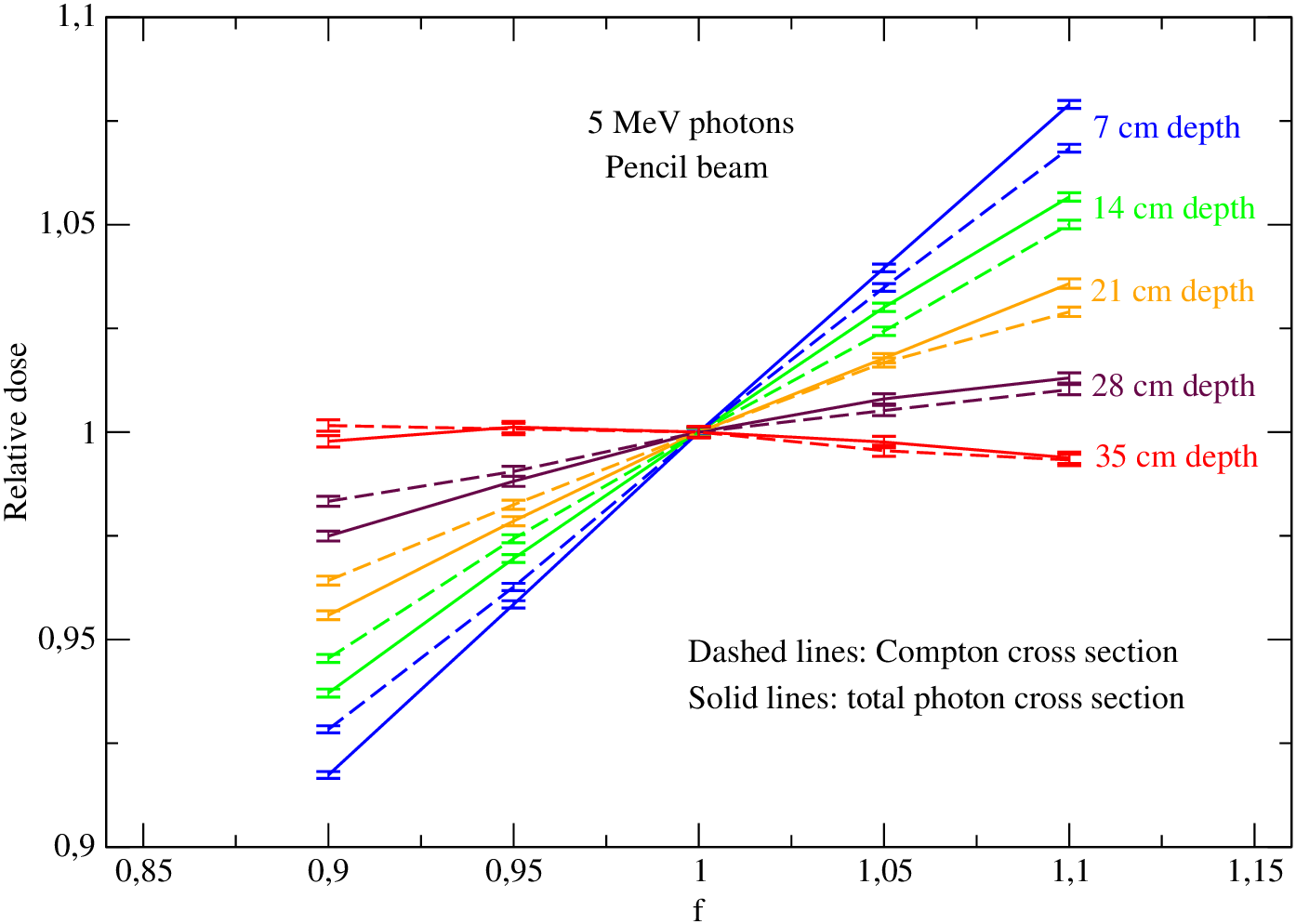}
        \caption{Compton}
        \label{fig_5MeV_a}
    \end{subfigure}
    \hspace{1cm}
    \begin{subfigure}[b]{0.45\textwidth}
        \includegraphics[width=\textwidth]{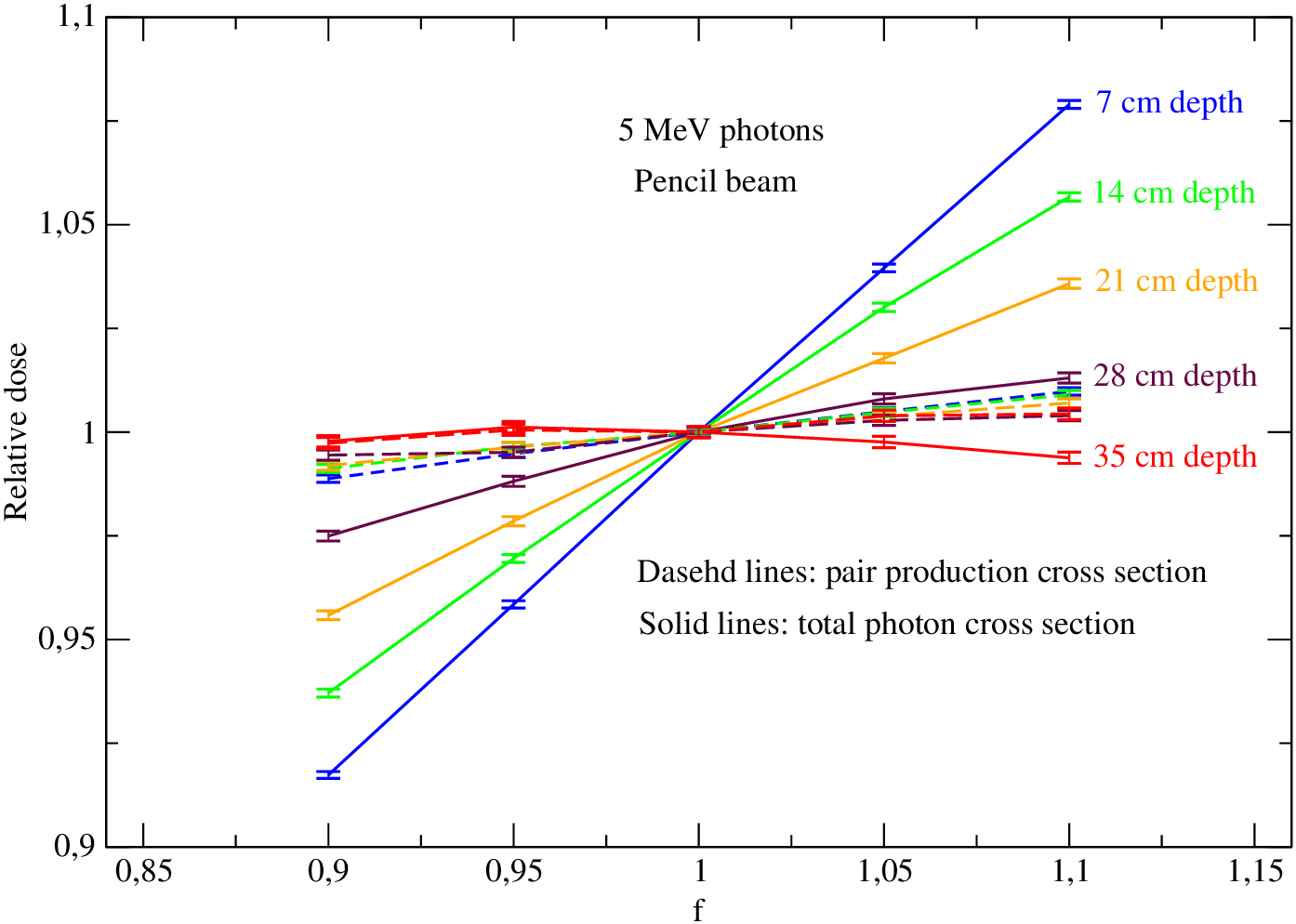}
         \caption{Pair production}
         \label{fig_5MeV_b}
    \end{subfigure}
  \caption{\label{fig_5MeV}Ratio between the central axis doses for a 5 MeV pencil beam with and without scaling of total cross sections as a function of the scaling factor $f$. Solid lines represent the situation in which total cross sections were scaled. Dashed lines represent the cases where (a) Compton and (b) pair-production CS were scaled. Values are reported for five different depths.}
\end{figure}

Higher incident photon energies were also simulated in order to  study the influence of the pair-production CS on dose uncertainty. First, a 2 MeV photon beam was studied, which is an energy similar to a 6 MV X-ray beam mean energy. Even at this relatively high energy, the Compton scattering process almost totally determines the overall effect of CS uncertainty on that of dose. In other words, the behavior observed for 2 MeV photons is similar to that shown for the $^{60}$Co case (see fig. \ref{fig_Co}). Thus, no plot will be shown for this energy. Then, it was decided to increase the incident energy up to 5 MeV, which would represent the mean energy of a 15 MV X-ray beam. Fig. \ref{fig_5MeV} shows the corresponding results. Even at this relatively high energy, the Compton scattering continues to have a great impact on dose distribution (see fig. \ref{fig_5MeV_a}). Fig. \ref{fig_5MeV} shows that the pair-production process has much less influence on dose distribution than Compton scattering. As obtained for lower energies,  the ED obtained for this energy is similar to the mean free path reported in the literature ($\sim$33 cm, \cite{NISTphotons}).

Table \ref{tab_sensitivity} summarizes the most important sensitivity factors for the four energies  and the interaction processes studied in this work. These factors can be combined with the corresponding CS relative uncertainty using eq. (\ref{dose_cs3}) to propagate this uncertainty to that of the  absorbed dose. For instance, according to the classical paper of Hubbell \cite{Hubbel99}, the uncertainty associated to the photo-electric cross section in the 5-100 keV interval is about 2 \%. Then, the relative uncertainty of the dose only due to this process for a 20 keV photon beam at 1 cm depth is about 0.5 \%.

\begin{table}[!htb]
\centering
\caption{Sensitivity factors $\partial R/\partial f$ as a function of depth for the four energies studied in this work. Only relevant values are shown for each energy.} \vspace*{0.3cm}
\begin{tabular}{cccccc} \hline
 & \multicolumn{5}{c}{20 keV beam} \\ \hline
 
Depth (cm) & 1 & 2 & 3 & 4 & 5 \\
\hline

Compton & -0.097 & -0.285 & -0.485 & -0.587 & -0.785 \\
Rayleigh & -0.036 & -0.107 & -0.237 & -0.263 & -0.376 \\
Photo-electric & 0.241 & -0.096 & -0.668 & -1.184 & -1.751 \\ 
\hline
\hline
 & \multicolumn{5}{c}{100 keV beam} \\ \hline
Depth (cm) & 3 & 6 & 9 & 12 & 15 \\
\hline
Compton & 0.442 & -0.045 & -0.516 & -0.940 & -1.468 \\ 
\hline
\hline
 & \multicolumn{5}{c}{1.25 MeV beam} \\ \hline
Depth (cm) & 5 & 10 & 15 & 20 & 25 \\
\hline
Compton & 0.708 & 0.394 & 0.084 & -0.243 & -0.497 \\  
\hline
\hline
 & \multicolumn{5}{c}{2 MeV beam} \\ \hline
Depth (cm) & 5 & 10 & 15 & 20 & 25 \\
\hline
Compton & 0.777 & 0.544 & 0.300 & 0.066 & -0.171 \\  \hline
\hline
 & \multicolumn{5}{c}{5 MeV beam} \\ \hline
Depth (cm) & 5 & 10 & 15 & 20 & 25 \\
\hline
Compton & 0.765 & 0.628 & 0.485 & 0.357 & 0.290 \\
Pair-production & 0.116 & 0.077 & 0.091 & 0.064 & 0.068 \\ \hline
\end{tabular}
\label{tab_sensitivity}
\end{table}

\begin{figure}
\includegraphics[width =.5\linewidth]{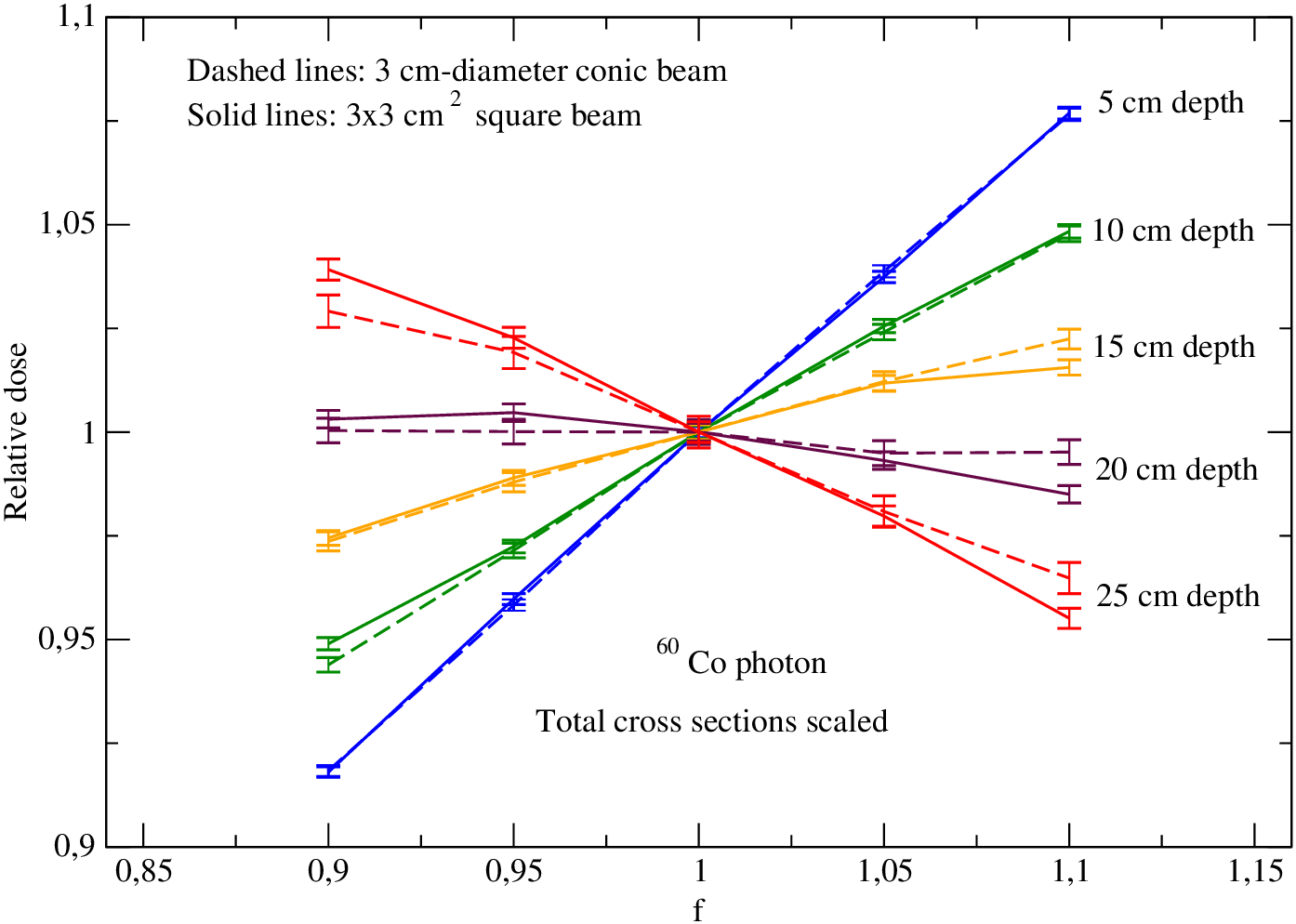}
\caption{Ratio between the central axis doses for a wide $^{60}$Co beam with and without scaling of total cross section as a function of the scaling factor $f$. Solid  and dashed lines represent the results for a normal 3x3 cm$^2$  square beam and a 3 cm-diameter conic beam, respectively.  Values are reported for five different depths.}
\label{fig_Co_field}
\end{figure}

The effect of the beam extension and divergence was also studied. For this purpose, only $^{60}$Co photons were studied as this is a very representative energy in Medical Radiology Physics. 3x3 cm$^2$ square and 3 cm-diameter conic fields were studied. The corresponding results  are shown in fig. \ref{fig_Co_field}.  The divergence of the conic beam is the same as that of a 10x10 cm$^2$ beam at a source-surface distance of 100 cm. The squared beam has no divergence.  The first important feature in this plot is that the influence of the field divergence increases with depth. This is a pure geometrical effect. In addition, the dose seems to be more sensitive to CS for the parallel field, at least for depths above the ED. The behavior for depths below the ED is rather undefined. The contribution to the central axis dose from a primary photon track in a wide beam depends on the photon scattering angular distribution and the relative angle between the incident ray and central axis. It is expected that this contribution is greater in a normal beam than in a divergent beam since the greater the angle between the incident ray and the beam central axis, the lower the contribution of this track to the dose on this axis. This difference increases with depth. Table \ref{tab_Co_field} shows the sensitivity factors for for these fields. If the results shown in tables \ref{tab_sensitivity} and \ref{tab_Co_field} for $^{60}$Co are compared, it can be concluded that the dose for extended fields is more sensitive to CS scaling than for pencil beam for depths below the ED. The opposite can be said for depths above the ED. We do not have a definite explanation for this behavior but it may be related to the scattering conditions in pencil and extended beams.

\begin{table}[h]
\caption{Square and conic $^{60}$Co beams sensitivity factors for total CS. The divergence of the conic beam is the same as that of a 10x10 cm$^2$ beam at a source-surface distance of 100 cm.}
\begin{center}
\begin{tabular}{ccc}
\hline
Depth (cm)	&	${\partial R/\partial f}$ Square beam & ${\partial R/\partial f}$ Conic beam \\
\hline
5			&	0.776							& 0.805	\\	
10			&      0.531							& 0.526   	\\
15			&      0.228							& 0.243	\\
20			&     -0.116							& -0.052	\\
25			&     -0.430							& -0.383  	\\
\hline
\end{tabular}
\end{center}
\label{tab_Co_field}
\end{table}%

 Finally,  table \ref{tab_overall} shows the combined dose uncertainty due to photon cross section systematic uncertainties  for the energies studies in this work as a function of depth. The energy and material independent photon CS uncertainty estimated by Ali et al. (0.5\% at 68\% confidence) \cite{Ali15} has been used in conjunction with eq. (\ref{propagation}) and the sensitivity factors shown in table \ref{tab_sensitivity}. It should be noticed that the combined uncertainty is negligible near the equilibrium depth, as commented above.
 
 \begin{table}[!htb]
\centering
\caption{Combined dose uncertainty due to photon CS unceratinties as a function of depth. The energy and material independent photon CS uncertainty of 0.5\% (68\% confidence) has been used \cite{Ali15}.} \vspace*{0.3cm}
\begin{tabular}{cccccc} \hline
 & \multicolumn{5}{c}{20 keV  beam} \\ \hline
 
Depth (cm) & 1 & 2 & 3 & 4 & 5 \\
\hline

 Combined uncer. (\%) & 0.13 & 0.16 & 0.43 & 0.67 & 0.98 \\
\hline
\hline
 & \multicolumn{5}{c}{100 keV beam} \\ \hline
Depth (cm) & 3 & 6 & 9 & 12 & 15 \\
\hline
 Combined uncer. (\%) & 0.22 & 0.02 & 0.26 & 0.47 & 0.73 \\ 
\hline
\hline
 & \multicolumn{5}{c}{1.25 MeV  beam} \\ \hline
Depth (cm) & 5 & 10 & 15 & 20 & 25 \\
\hline
 Combined uncer. (\%)  & 0.35 & 0.20 & 0.04 & 0.12 & 0.25 \\  
\hline
\hline
 & \multicolumn{5}{c}{2 MeV  beam} \\ \hline
Depth (cm) & 5 & 10 & 15 & 20 & 25 \\
\hline
  Combined uncer. (\%) & 0.39 & 0.27 & 0.15 & 0.03 & 0.09 \\  \hline
\hline
 & \multicolumn{5}{c}{5 MeV  beam} \\ \hline
Depth (cm) & 5 & 10 & 15 & 20 & 25 \\
\hline
 Combined uncer. (\%) & 0.39 & 0.32 & 0.25 & 0.18 & 0.15 \\
 \hline
\end{tabular}
\label{tab_overall}
\end{table}

\section{Conclusions}
 The sensitivity of the central axis dose to changes in photon cross sections depends on the depth into the phantom. This sensitivity is nearly constant for photon CS uncertainties variations of about $\pm$10\%. For photon energies of the order of a few tens of keV, the dose uncertainty is sensitive to the uncertainty of photo-electric effect and Compton and Rayleigh scattering cross sections. From approximately 100 keV to about 2 MeV, this sensitivity is only seen for the Compton scattering cross section. Above about a few MeV, the dose uncertainty begins to depend on the pair-production cross section uncertainty, besides that for Compton scattering.
 
 There is an equilibrium depth at which sensitivity factors are negligible  and so the impact of CS uncertainties on the dose uncertainty. This depth is practically determined by the photon mean free path for the incident photon energy. On the one hand, the  underestimation of photon CS produces a decrease and increase of the dose at depths below  and above the equilibrium depth, respectively. On the other hand, the  overestimation of photon CS produces an increase and decrease of the dose at depths below  and above the equilibrium depth, respectively. The impact of photon CS uncertainties on depth-dose curves determined by Monte Carlo simulations depends on depth. This is an important issue  when using MC-based treatment planning systems.  The impact of electron interaction cross sections can be studied in a future, having into account that electron transport algorithms are  code-dependent. It should be a much more difficult task.

\section*{Acknowledgment}
M. Bernal thanks the FAPESP foundation in S\~ao Paulo, Brazil, for financing his research activities through the 2011/51594-2 project. E. Aguirre acknowledges the financial support from the Coordena\c{c}\~ao de Aperfei\c{c}oamento de Pessoal de N\'ivel Superior (CAPES) in Brazil.

\section*{References}

\end{document}